\documentclass[]{spie}  

\usepackage{amsmath,amsfonts,amssymb}
\usepackage{graphicx}
\usepackage[colorlinks=true, allcolors=blue]{hyperref}
\usepackage[utf8]{inputenc}

\title{The InfraRed Imaging Spectrograph (IRIS) for TMT: latest science cases and simulations}

\author[a,b]{Shelley A. Wright}
\author[a,b]{Gregory Walth}
\author[c]{Tuan Do}
\author[a,b]{Daniel Marshall}
\author[c]{James E. Larkin}
\author[d]{Anna M. Moore}
\author[e]{Mate Adamkovics}
\author[f]{David Andersen}
\author[g]{Lee Armus}
\author[h]{Aaron Barth}
\author[f]{Patrick Cote}
\author[i]{Jeff Cooke}
\author[j]{Eric M. Chisholm}
\author[f]{Timothy Davidge}
\author[f]{Jennifer S. Dunn}
\author[j]{Christophe Dumas}
\author[j]{Brent L. Ellerbroeck}
\author[c]{Andrea M. Ghez}
\author[k]{Lei Hao}
\author[l]{Yutaka Hayano}
\author[m]{Michael Liu}
\author[n,o]{Enrique Lopez-Rodriguez}
\author[m]{Jessica R. Lu}
\author[p]{Shude Mao}
\author[f]{Christian Marois}
\author[q]{Shashi B. Pandey}
\author[r]{Andrew C. Philips}
\author[f,j]{Matthias Schoeck}
\author[s]{Annapurni Subramaniam}
\author[t,u]{Smitha Subramanian}
\author[l]{Ryuji Suzuki}
\author[v]{Jonathan C. Tan}
\author[l]{Tsuyoshi Terai}
\author[c]{Tommaso Treu}
\author[f]{Luc Simard}
\author[c]{Jason L. Weiss}
\author[d]{James Wincensten}
\author[e]{Michael Wong}
\author[w]{Kai Zhang}

\affil[a]{Department of Physics, University of California San Diego, CA, 92039, USA;}
\affil[b]{Center for Astrophysics and Space Sciences, University of California San Diego, CA, 92039, USA;}
\affil[c]{Physics and Astronomy Department, University of California Los Angeles, CA 90095 USA;}
\affil[d]{Caltech Optical Observatories,1200 E California Blvd., Pasadena, CA 91125 USA;}
\affil[e]{Astronomy Department, University of California Berkeley, CA 94720 USA;}
\affil[f]{National Research Council of Canada - Herzberg, Victoria, BC, V9E 2E7 Canada;}
\affil[g]{Infrared Processing and Analysis Center, California Institute for Technology, Pasadena, CA 91125, USA;}
\affil[h]{Department of Physics and Astronomy, University of California Irvine, Irvine, CA 92697 USA;}
\affil[i]{Centre for Astrophysics and Supercomputing, Swinburne University of Technology, Hawthorn, VIC 3122, Australia;}
\affil[j]{Thirty Meter Telescope Observatory Corporation, Pasadena, CA 91105 USA;}
\affil[k]{Shanghai Astronomical Observatory, Shanghai 200030, China;}
\affil[l]{National Astronomical Observatory of Japan, Osawa, Mitaka, Tokyo, 181-8588 Japan;}
\affil[m]{Institute for Astronomy, University of Hawaii, Manoa, HI 96822 USA;}
\affil[n]{Department of Astronomy, University of Texas at Austin, TX, 78712, USA;}
\affil[o]{McDonald Observatory, University of Texas at Austin, TX, 78712, USA;}
\affil[p]{National Astronomical Observatories, Chinese Academy of Sciences, Beijing 100012, China;}
\affil[q]{Aryabhatta Research Institute of Observational Sciences, Manora Peak, Nainital, Uttarakhand, India;}
\affil[r]{University of California Observatories, Santa Cruz, CA USA;}
\affil[s]{Indian Institute of Astrophysics II Block Koramangala, Bangalore 560 034 India;}
\affil[t]{Kavli Institute for Astronomy and Astrophysics, Peking University, Hai Dian District, Beijing 100871, China,;}
\affil[u]{Department of Astronomy, Peking University, Hai Dian District, Beijing 100871, China;}
\affil[v]{Departments of Astronomy and Physics, University of Florida, Gainesville, Florida 32611, USA;}
\affil[w]{National Astronomical Observatories, Nanjing Institute of Astronomical Optics and Technology, Jiangsu District, China}

\authorinfo{Send correspondence to S. Wright: saw@physics.ucsd.edu}

\begin{document} 
\maketitle

\begin{abstract}
The Thirty Meter Telescope (TMT) first light instrument IRIS (Infrared Imaging Spectrograph) will complete its preliminary design phase in 2016. The IRIS instrument design includes a near-infrared (0.85 - 2.4 micron) integral field spectrograph (IFS) and imager that are able to conduct simultaneous diffraction-limited observations behind the advanced adaptive optics system NFIRAOS. The IRIS science cases have continued to be developed and new science studies have been investigated to aid in technical performance and design requirements. In this development phase, the IRIS science team has paid particular attention to the selection of filters, gratings, sensitivities of the entire system, and science cases that will benefit from the parallel mode of the IFS and imaging camera. We present new science cases for IRIS using the latest end-to-end data simulator on the following topics: Solar System bodies, the Galactic center, active galactic nuclei (AGN), and distant gravitationally-lensed galaxies. We then briefly discuss the necessity of an advanced data management system and data reduction pipeline.


\end{abstract}


\section{INTRODUCTION}\label{sec:intro}  

While ground-based giant-segemented telescope (GSMTs) - like the European
Extremely Large Telecscopes (E-ELT)\cite{Ramsay}, Giant Magellan Telescope
(GMT)\cite{Jacoby}, and the Thirty Meter Telescope (TMT)\cite{Skidmore} -
are still several years away from completion, important groundwork is being
laid right now to build a strong foundation for their first light programs.
The scientific capabilities of GSMTs will both complement and extend the
astronomical discovery space of upcoming facilities like the James Webb
Space Telescope (JWST)\cite{Greenhouse}, the Wide Field Infrared Survey
Telescope (WFIRST)\cite{Spergel}, and the Large Synoptic Survey Telescope
(LSST)\cite{Gressler}. For instance, spectroscopic follow-up from GSMT
instruments will be essential to the LSST mission. Additionally, when
compared with NIRspec\cite{Birkmann} integral field spectrograph and
NIRcam\cite{Beichman} on JWST, instruments like IRIS (InfraRed Imaging
Spectrograph)\cite{Larkin2016} for TMT will be able to provide 25 times
more angular resolution, and for spectroscopy 2-3 times higher spectral
resolving power.  The diffraction-limit of a 20m telescope or greater
aperture affords scientific potential that is unique to {\it all} current and future facilities, and will address areas that are fundamental to our understanding of the universe. 

IRIS is the first-light instrument being designed to operate with the advanced adaptive optics (AO) system, NFIRAOS\cite{Glen}, for the TMT. IRIS science instruments include a near-infrared (0.84 - 2.45 micron) integral field spectrograph (IFS) and imaging camera, as described in greater detail in Larkin et al.\cite{Larkin2016}. IRIS with NFIRAOS will make use of three on-instrument wavefront sensors (OIWFS)\cite{Dunn} to provide focus and tip/tilt correction. During the current preliminary design phase (2015-2016) of IRIS, the technical team has developed a new optical design that allows the IFS and imaging camera to share common optics and provide a unique parallel observing mode, where the IFS and imaging camera operate simultaneously. 

We have assembled a large international science team from six countries to define IRIS’s capabilities, requirements, and science case studies. Our team has investigated many science cases that IRIS will offer to a range of astronomical fields from the solar system to first light galaxies\cite{Barton10,Wright14}. An end-to-end simulator for both the IRIS IFS and imaging camera has been developed to explore point source and resolved source sensitivities for particular use cases\cite{Wright10,Do14}. This has been helpful for exploring astrometric and photometric capabilities for IRIS and other technical requirements\cite{astrometry}. For instance, this software was used to perform a detailed study of IRIS's potential for measuring the masses of supermassive black holes (SMBH)\cite{Do14}, and to explore potential measurements of first light galaxies (z $\gtrsim$ 9). Our team has continued to investigate new IRIS science cases.  In this paper, we present the latest simulations for IRIS.  Section 3.1 describes monitoring solar system bodies, while Section 3.2 discusses parallel imaging and spectroscopy of the Galactic Center. In Section 3.3, we discuss the possibility of resolving an AGN torus (Section 3.3), and describe detailed study of gravitationally-lensed distant galaxies in Section 3.4. These are just a representative sample of IRIS and TMT science cases that highlight the uniqueness of combining these diffraction-limited data sets.


\section{DATA SIMULATOR}
\label{sec:sim}  
The IRIS team has developed a full end-to-end data simulator for both the imager and integral field spectrograph\cite{Wright10,Do14}. The simulator has been used to determine IRIS sensitivities, explore a range of science cases, and help to investigate instrumental requirements. We have previously described the technical details for the simulator and presented direct point source and resolved source sensitivities. Briefly, the simulator generates the predicted signal-to-noise (S/N) of an image or IFS data cube assuming realistic noise, a background model, and point spread functions (PSF) from TMT and NFIRAOS. From a user perspective, the simulator needs an input of a flux calibrated image, a spectrum for the IFS, and observing parameters (e.g., filter, plate scale, exposure time, number of frames).

\section{SELECTION OF IRIS SCIENCE CASES}
\label{sec:cases} 
    
\subsection{Solar System}

IRIS will be tremendously advantageous for a number of Solar System science studies, ranging from near earth asteroids (NEAs), planets, dwarf planets, gas giant moons, and transneptunian objects (TNOs). Using the data simulator, we have explored two new IRIS Solar System cases: the Pluto/Charon system and Jupiter's active moon, Io.

\begin{figure*}
  \centering
  \includegraphics[width=.9\linewidth]{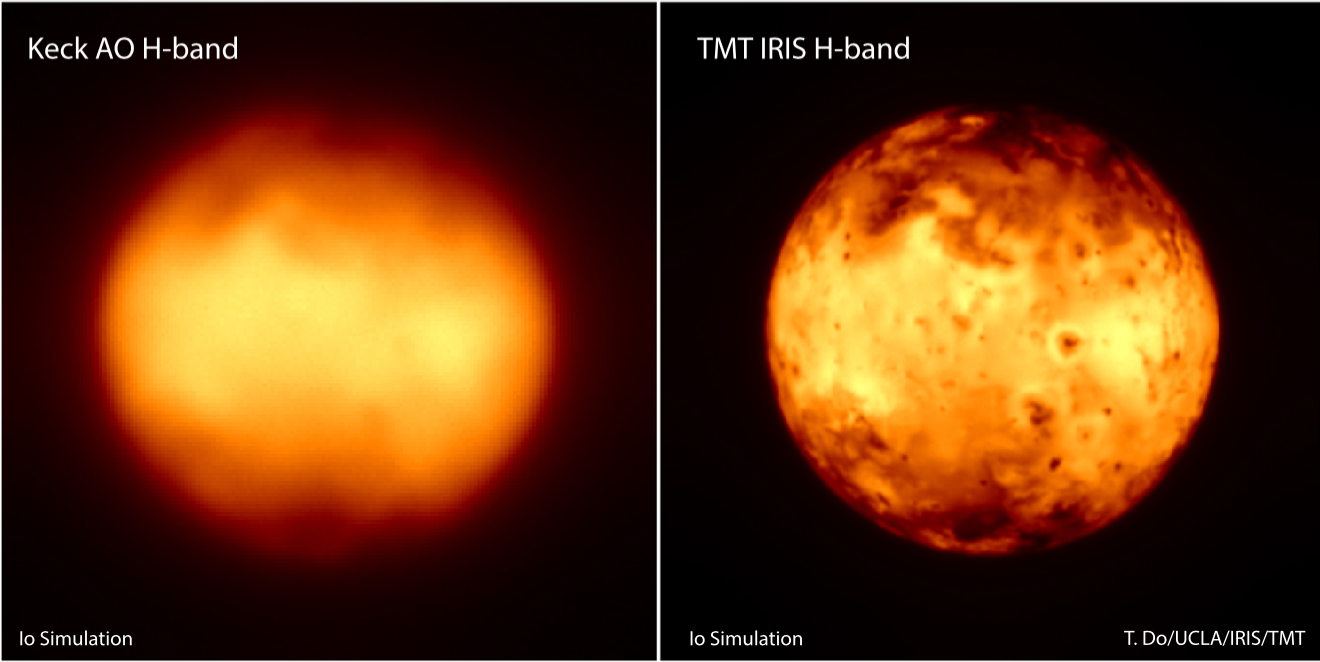}
  \caption{Keck AO/NIRC2 simulation image compared to simulated IRIS image of Io. Both Simulations made use of the Galileo Solid-State imaging camera data set. IRIS will be able to sample down to $\sim$12 km using the 4 mas plate scale of the imager and IFS. The typical size scales of volcanos on Io's surface range from 40 to 200 km.}
  \label{fig:io}
\end{figure*} 

\subsubsection{Io}
\label{sec:Io}  

Io has the most active and volatile surfaces in our Solar System due to strong-tidal interactions with Jupiter. Planetary scientists and geologist have great interest in Io's frequent violent volcanic eruptions, with hundreds of identified volcanic sites. Even though Io has been studied in detailed by spacecraft missions like \textit{Voyager} and \textit{Galileo} and other telescopes, there are still unanswered questions about the eruption characteristics like the frequency, temperatures, lava composition, and eruption type that require regular monitoring of Io's surface. Currently, ground-based integral field spectrographs and AO systems have been an excellent resource for discovering new volcanic regions and investigating the compositions of these eruptions\cite{Marchis,Laver07,Kleer,Pater16}. 

With the IRIS data simulator and Galileo imaging at 1$\mu$m, we simulated an IRIS H-band image of Io using a PSF calculated near zenith and median conditions. The Galileo image, taken from the Solid-State-Imaging camera\footnote{http://photojournal.jpl.nasa.gov/catalog/PIA00494}, was scaled to a total flux of 5.8 mag in H-band with an exposure time of only one second.  In Figure \ref{fig:io}, we compare these simulated observations to the current capabilities of the Keck AO system with NIRC2 at H-band.  The exquisite resolution of IRIS+TMT is exemplified in this comparison, showing that the instrument will be excellent for probing Solar System bodies in detail.  IRIS will resolve structures on Io as small as $\sim$20 km, only a factor of 2 larger than the resolution of \textit{Galileo}.

\subsubsection{Pluto and Charon}
\label{sec:pluto}  

The \textit{New Horizons}\cite{Stern} spacecraft was able to image the surface of the Pluto at unprecedented detail, and conduct essential spectroscopic observations of its atmosphere. From this mission, we learned that Pluto has an active surface\cite{hammond} and atmosphere, with volatile ices\cite{Grundy} and sheets across the surface, as well as volatile atmospheric gases\cite{Gladstone}. There are also seasonal and global latitude weather changes observed on Pluto that need further monitoring on monthly and yearly timescales. The Pluto/Charon system has an additional four small moons, and the combination of HST and New Horizons data has been excellent for investigating their dynamics and stability\cite{Showalter}. With IRIS, we will be able to monitor all of these bodies with high astrometric accuracy.

\begin{figure*}
  \centering
  \includegraphics[width=.9\linewidth]{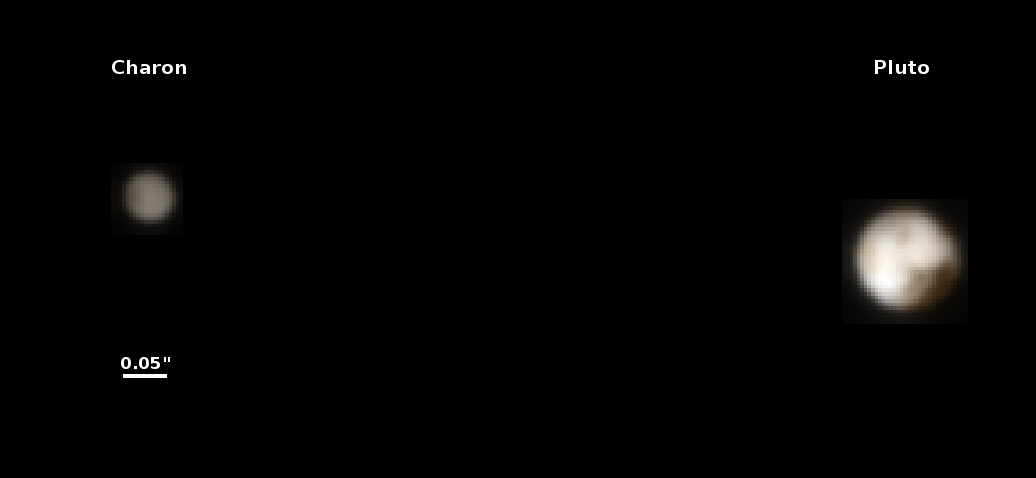}
  \caption{Simulated IRIS three color ($J$, $H$, and $K$) image of Pluto and Charon, assuming a \textit{single} 100 second integration time with \textit{no} deconvolution performed. In comparison, HST observations only a sample $\sim$10s pixels across Pluto and need to perform extensive deconvolution routines on multiple phases of observations\cite{Buie}. In this single observation, the IRIS imager at the 4 mas scale will resolve a spatial scale of $\sim$82 km on both Pluto and Charon.}
  \label{fig:pluto}
\end{figure*}

Using the highest resolution images of Pluto and Charon from the \textit{New Horizons} spacecraft, we simulated IRIS performance and capabilities. We made use of optical images from the Long Range Reconnaissance Imager (LORRI)\cite{Conard}, and scaled the optical bands based on the spectral shape of Pluto and Charon to the near-infrared passbands ($J$, $H$, and $K$). Figure \ref{fig:pluto} is a three color composite simulation with the IRIS imager (4mas plate scale).  We find that IRIS easily resolves features on both Pluto and Charon. Each resolution element will be able to resolve $\sim$100km regions across the surface of Pluto and take high-resolution spectroscopy. These type of studies will be superb for monitoring seasonal and atmospheric changes across the globe.  In addition, IRIS will be used to monitor Pluto and Charon's surface with the IFS at relatively high resolution spectroscopy (R=4000, 8000). The diffraction-limit of TMT and IRIS will offer $>$500 IFS sampling points across Pluto's surface, as seen in Figure \ref{fig:pluto}, which is an order magnitude higher than all previous HST and future JWST IFS observations.

\subsection{Galactic Center}
\label{sec:GC}  

\begin{figure*}
  \centering
  \includegraphics[width=.9\linewidth]{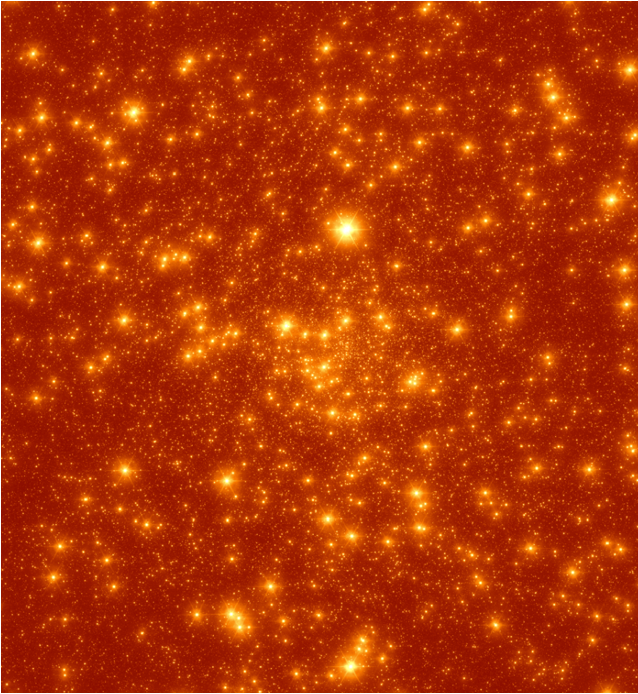}
  \caption{Simulation of an IRIS imager observation using K (2.2 $\mu$m) broadband of the Galactic Center with the expanded field of view of 32.8” x 32.8”. In a single shot using the 0.004” plate scale the IRIS imager will be able to capture the inner-arcsecond sources, Sgr A*, the extended stellar population, and essential astrometric reference frame sources of 8 masers. This simulated observation represents only 20 seconds of exposure time. Based on the extrapolated stellar population IRIS would be able to detect 500,000 stars down to K $<$ 25 mags.}
  \label{fig:gcim}
\end{figure*}

The Galactic Center is the closest laboratory for studying the environments and fundamental physics of SMBH\cite{ghez} and offers very unique science cases for TMT. Both the IRIS imager and IFS have been carefully designed to characterize the surroundings of the SMBH, SgrA*, at the Milky Way’s center. The high relative astrometric accuracy of 30 $\mu$as will offer a test of General Relativity and probe the distribution of dark matter through orbital monitoring of the stars surrounding SgrA*\cite{Yelda14}. The estimated astrometric accuracy is based on a reference frame that makes use of several radio maser sources with accurate positions in the Galactic Center. Currently, the limited field of view of Keck and VLT observations means that the observations have to be dithered to capture all of these masers. These dithers and the optical distortion severely limit the astrometric accuracy achieved at the Galactic Center today. Using the IRIS expanded imager capability ($\sim$34"x34") we will be able to observe 8 maser sources in a single observation, which will greatly improve the astrometric accuracy in this field. This is illustrated in Figure \ref{fig:gcim} that shows an IRIS simulation of a single K-band integration on the Galactic Center at median conditions and airmass from Maunakea. The improved astrometry is essential for measuring the orbits of current and new inner-arcsecond sources surrounding Sgr A* and exploring the fundamental physics of the SMBH.

IRIS IFS will yield greater accessibility and ease of studying the stellar population at the Galactic Center. The young stellar population near SgrA* has puzzled astronomers, as young massive stars should have difficulty forming in close proximity to a SMBH. Researchers have thus far been limited to studying only the most luminous stars in the area, including OB main sequences stars, red giants, and Wolf-Rayet stars. Currently, OSIRIS on Keck is able to achieve spectroscopy with sufficient SNR to measure spectral types and radial velocities for $Kp$ $<$ 15.5 mag stars\cite{tuan13}. In contrast, IRIS is predicted to have the sensitivity to allow for high SNR spectroscopy on $Kp$ = 20 - 21 mag stars. These sensitivities and high angular resolution will allow researchers to study the low mass-end of the main sequence, which will be crucial for investigating the stellar population. They will also provide essential radial velocity measurements to couple with the proper motion monitoring via imaging for deriving 3D orbital solutions.

\begin{figure*}
  \centering
  \includegraphics[width=.7\linewidth]{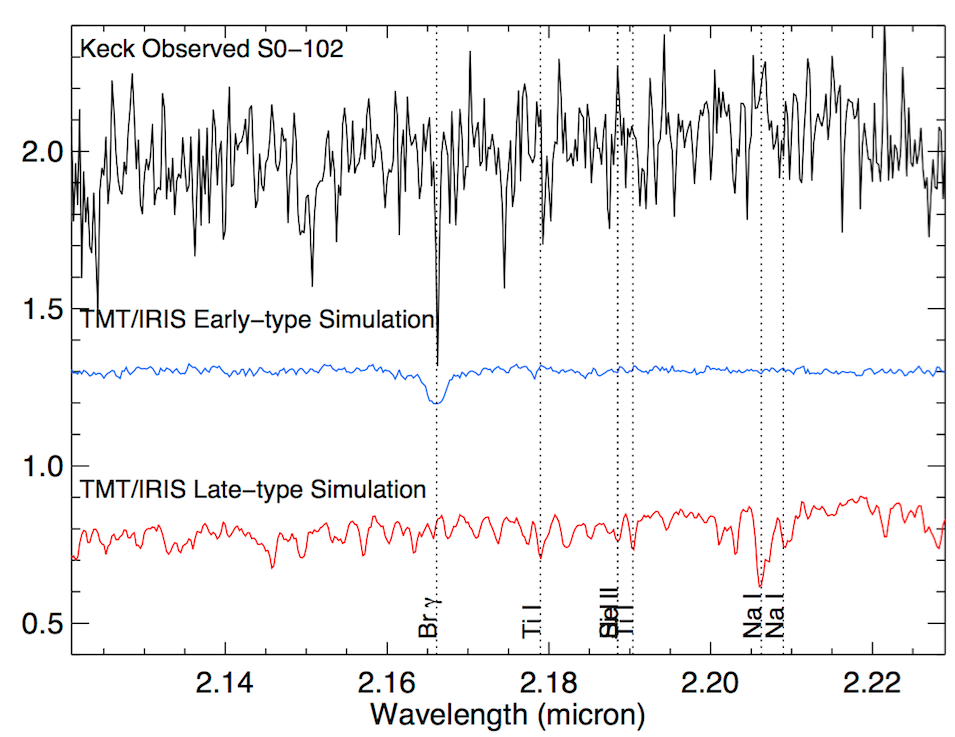}
  \caption{Comparison between 6.75 hours of observations from Keck on a Galactic Center source to only 15 minutes of an observation from the IRIS simulator. The inner-arcsecond source, S0-102, is simulated using the $Kn3$ filter with the 4mas plate scale. We assume that S0-102 has K=17.5 mag. The IRIS simulation assumes an total integration time of 900 seconds with a PSF at Zenith angles observable from Mauna Kea with median conditions. Since the current Keck observations are unable to distinguish the spectral type of S0-102, we simulate both an early-type and late-type star for IRIS+TMT. The IRIS simulation shows that we can achieve an integrated S/N of 270. In contrast, the Keck data has a 900 second integration with 27 frames and achieves a S/N of 27.  TMT+IRIS which will easily measure the spectral type of the source and distinguish between multiple stellar population models. These simulations highlight the unique capabilities that the IRIS IFS will have on the Galactic Center sources for stellar population studies and radial velocity measurements.}
  \label{fig:gcspec}
\end{figure*}

The shortest period star (S0-102) identified to-date among the inner-arcseond sources surrounding SgrA* has a period of 11.5 years\cite{Meyer12}. IRIS is expected to identify multiple sources with shorter orbital periods (1-2 years), which will be instrumental for fundamental physics studies like testing General Relativity. One of the current challenges with Keck and VLT spectroscopy is identifying the accurate spectral types for faint Galactic Center sources. For instance, S0-102's current spectral type is ambiguous between either an early-type or late-type star due to insufficient SNR. In Figure \ref{fig:gcspec}, we show a comparison between IRIS simulations using synthetic spectral models of early- and late-type stars at $K$-band to the latest Keck observations of S0-102. The resolution and sensitivity of the IRIS IFS will be capable of easily distinguishing between absorption features like Br$\gamma$ and Na-I. Resolving these absorption line features will also be crucial for measuring accurate radial velocities, which will be included in the full 3D orbital solution.

\subsection{Nearby Galaxies}
\label{sec:GC}  

There are a diverse number of nearby galaxies that can be studied using the IRIS imager and IFS. The IRIS science team has focused on variety of science cases, including: measuring SMBH masses; studies of galactic nuclei and AGN; chemical enrichment histories of galaxy types; metal enrichment of galaxy clusters; dynamics of dwarf galaxies; and probing luminous and ultra-luminous infrared galaxies. The IRIS diffraction-limited capabilities will allow exquisite stellar population studies in galaxies to distances as far as the Virgo cluster (16.5 Mpc), and will be able to resolve individual stars in these systems. Similarly, IRIS high angular resolution offers the exciting possibility of resolving and conducting dynamical studies of galactic nuclei and of gas and stars surrounding AGN. Recently, our team explored IRIS capabilities of resolving an AGN torus at near-infrared wavelengths.

\subsubsection{Resolving AGN torus}
\label{sec:AGN}

The unified model of Active Galactic Nuclei (AGN) posits that different observational classes of AGNs can be explained by an orientation effect \cite{Antonucci}. Specifically, it is thought that differences between radio-quiet AGN classifications are due to different orientations of an optically thick, torus shaped accumulation of dust which obscures the AGN's central black hole and accretion disc. Current telescopes have not been able to directly resolve this proposed dusty torus, a feat which could help determine the extent to which orientation effects are responsible for differences between AGN classifications. 

\begin{figure}
  \centering
  \includegraphics[width=.9\linewidth]{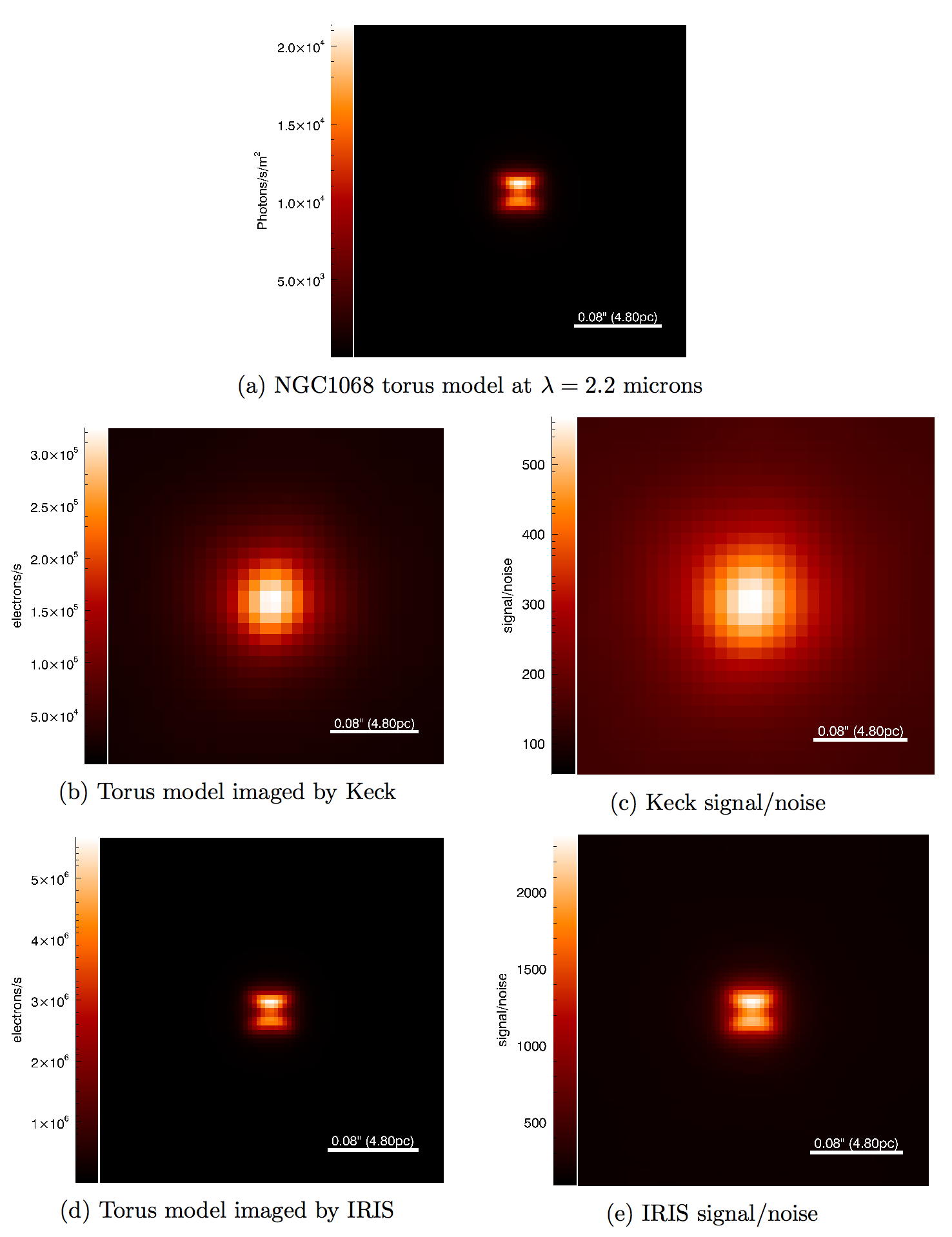}
  \caption{(a) Model of NGC1068's torus region ($\lambda = 2.2$ microns, 0.004" per pixel), an archetypal Type II Seyfert galaxy. (b) Simulated Keck observation (K band, 0.01" per pixel, 1 second exposure) of the model with the corresponding SNR shown in (c). (d) Simulated IRIS observation (K band, 0.004" per pixel, 1 second exposure) of the model with the corresponding SNR shown in (e).  The IRIS simulated images clearly resolve the structure and orientation of the simulated torus, something that is not currently possible given the Keck simulations.}
  \label{fig:torus}
\end{figure}

To determine whether IRIS will be able to resolve an AGN torus, we constructed a model for the torus of NGC1068 (the brightest Type II Seyfert galaxy) to use in the IRIS simulator.  We assumed that the three major sources of flux at $\lambda = 2.2$ microns originating from the torus region of NGC1068 are the accretion disk, the torus itself, and diffuse stellar emission. But a VLT interferometry study estimated that the contribution of the accretion disc emission to the total flux at $\lambda = 2.0$ microns is negligible compared to the torus emission\cite{Honig}. Therefore, our torus model (see Figure \ref{fig:torus}a) is composed of the following two components: a S{\'e}rsic profile (for the diffuse stellar emission) combined with a CLUMPY torus model (clumpy.org).

The geometric parameters of the S{\'e}rsic profile were obtained via fitting to the HST F222M observation of NGC1068. The CLUMPY torus model was generated assuming a distance of 12.5 Mpc, with an angular scale of 0.0167"/pc. The fixed torus parameters were assumed to be the MAP values of the Bayesian Clumpy fitting\cite{Alonso}. 
The variable parameters $Y$ and $q$, where $Y$ is the extension of the torus (estimated as the ratio of the inner and outer edge of the torus) and $q$ is the index of the radial distribution of the clouds, were chosen to be $Y=5$ and $q=2$, representing a small and compact torus. The relative contributions to the total flux from each component (i.e., the torus and the S{\'e}rsic model) was scaled such that, in a 0.5 arcsec aperture centered on the AGN, 55\% of the total flux originates from the torus and 45\% originates from diffuse stellar emission - these percentages were deduced through analysis of MMT-POL observations of NGC1068 in the $K^\prime$ filter ($\lambda_c = 2.20$ microns, $\Delta \lambda_{fwhm} = 0.08$ microns)\cite{Rodriguez}. 

Simulated observations of the modeled torus using both IRIS and Keck were generated. Figure \ref{fig:torus}b shows the image detected by Keck (for 1 sec of integration time in the K-band) and Figure \ref{fig:torus}d shows the image detected by IRIS (also for 1 sec of integration time in the $K$-band). Figures \ref{fig:torus}c and \ref{fig:torus}e show the corresponding signal to noise ratios for each pixel obtained by considering the noise from the detector and sky.  We find that IRIS will have sufficient contrast and resolution to resolve the AGN torus of NGC1068, an exciting possibility for confirming the unified AGN model.  Our simulations also imply that IRIS will have the ability to resolve torii for other nearby Seyfert systems, opening up a new and exciting science opportunities. This will be incredibly powerful when utilizing the IFS at R=4000 and even higher at R=8000 for characterizing the dynamics of the torus and spectral characterizations.

\subsection{Distant Galaxies}
\label{sec:highz}  

IRIS will be uniquely positioned to conduct spatially resolved near-infrared spectroscopy of distant (z $>$1) young galaxies on a range of science topics. Resolved spectroscopy will be essential for spatially resolving rest-frame ultraviolet emission lines, like, Ly$\alpha$ and H$\alpha$, from star-forming regions in the galaxies and/or any potential AGN, as highlighted in our team's previous simulations\cite{Wright14}. IRIS will also be superb for conducting resolved stellar population studies of distant galaxies. This will be critical for combining the resolved dynamics of stars with the nebular gas observations at giant molecular cloud resolutions. Combining this with the magnification gain of gravitationally-lensed systems will be even more powerful, and the finer plate scales of IRIS will be uniquely positioned compared to other facilities like JWST. 

Exploring transients in distant galaxies is an exciting area of exploration in which IRIS will play a roll. Identifying and studying the local environments of supernovae (SNe) and transients in z $>$ 2 galaxies will be essential for understanding their progenitors and their connection to galaxy properties. The IFS will be well-suited for identifying the spectral signatures in distant SNe, superluminous supernovae (SLSNe)\cite{Cooke,Tanaka}, and potentially optical counterparts to elusive fast radio bursts\cite{Spitler}. IRIS's sensitivity will also provide important constraints on the properties of distant Gamma Ray Bursts, using them to probe the intervening inter-galactic medium.

The parallel mode of IRIS will provide extraordinary data sets on deep extragalactic fields. While spectroscopic observations are being made on a particular high-redshift galaxy (z $>$ 1), the imager is now designed to be able to take simultaneous observations.  With this mode, IRIS imager  be able to achieve deep observations within 20 minutes that are comparable to the Hubble Deep Fields. This level of multiplexing will be outstanding for galaxy studies at all distance scales, as further described in Larkin et al.\cite{Larkin2016}.

\subsubsection{Resolved Spectroscopy and Imaging of Gravitationally-Lensed Galaxies}
\label{sec:highz} 

Strong gravitational lensing provides a powerful way of studying high-redshift galaxies at high angular resolution with greater sensitivity to probe down to fainter galaxies and lower-surface brightness regions. Current Keck and VLT AO studies of high-redshift galaxies are limited to the brightest and most massive star forming galaxies, with the typical angular resolution of $\sim$1 kpc\cite{Glazebrook}. Lensed galaxies with large magnifications can an get an increased resolution by 5-20x and when combined with IRIS+TMT it will open a new window for studying distant galaxies at even Giant Molecular Cloud size scale (\textit{10s of parsecs}) enabling direct comparisons to galaxies in the local Universe. Even using ALMA at its longest baselines will only provide 20 mas resolution, which is 5$\times$ larger than the highest resolution achievable by IRIS+TMT. 

\begin{figure*}
  \centering
  \includegraphics[width=1.\linewidth]{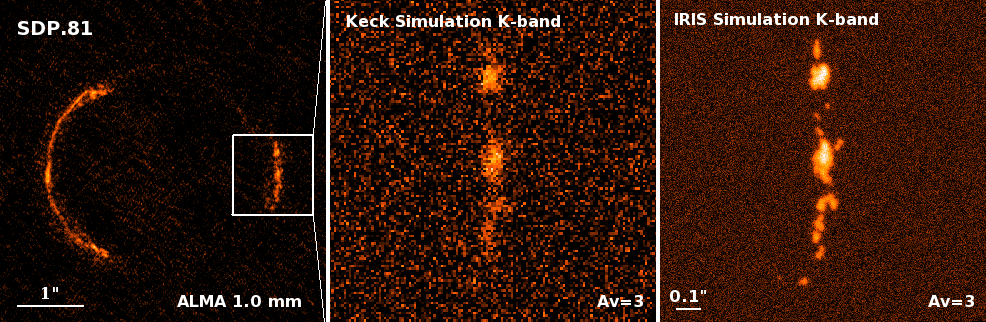}
  \caption{IRIS simulation of a distant strong gravitationally-lensed galaxy. (LEFT) Original ALMA 1.0 mm image of the gravitationally-lensed galaxy SDP.81 at z=3.042 with 23 mas resolution used for Keck/NIRC2 and TMT/IRIS simulations. (MIDDLE) A zoom-in of Keck/NIRC2 K-band simulated image of the arc assuming that the stellar continuum is attenuated by a Calzetti dust law with an A$_{V}$=3. (RIGHT) IRIS imager K-band simulation using the same parameters as Keck/NIRC2. The 23 mas structures are clearly resolved with IRIS, and at the highest spatial resolution it will be able to resolve structures $\sim$5$\times$ smaller than ALMA. Both simulations were performed with the same total integration time of 8 hours for a direct comparison.}
  \label{fig:sdp81}
\end{figure*}

Using one of the highest resolution images ever taken by ALMA of a gravitationally-lensed galaxy (SDP.81\cite{ALMA_SDP81_2015}), we generated an IRIS simulation of its expected stellar continuum emission. SDP.81 was imaged by ALMA for the science verification campaign\cite{ALMA2015} of their longest baseline ($\sim$15 km). SDP.81\cite{Negrello2010} is a gravitationally-lensed, dusty, star-forming galaxy at z=3.042. The lens is a galaxy at z=0.2999. The background galaxy is almost perfectly aligned behind the lens such that it produces an Einstein ring. With this lensing configuration, SDP.81 is magnified by a factor of 11. At the 15 km baseline, ALMA's spatial resolution is $\sim$23 mas or 180 pc at z=3.042. Taking into account the lens model, the source plane is able to sample at $\sim$10 pc resolution.

To generate a simulation image for both Keck NIRC2 and TMT IRIS, we first apply a 5$\sigma$ threshold on the ALMA 1.0 mm image in order to ensure that we are including real features for our simulations. We then assume the stellar continuum is spatially coincident with the dust emission. This assumption may not be true, but it allows us to simulate a representative stellar surface brightness profile of the galaxy. We then fit spectral energy distributions to Herschel (250, 350 and 500 $\mu$m), SMA\cite{Bussmann2013} (880 $\mu$m) and MAMBO (1.2 mm) data using SDP.81 templates\cite{CharyElbaz2001}. Next, we estimate the stellar continuum from the SED template redshifted into K-band. Finally, using a Calzetti\cite{Calzetti2001} dust attenuation law, we apply a range of dust attenuations (A$_{V}$=0-7) for a diverse set of IRIS simulations.  Each of the simulations are run with 300 second integrations with 100 frame stacks ($\sim$8 hours on source), shown in central and right panels of Figure~\ref{fig:sdp81}. We choose an 8 hour total integration, so both Keck/NIRC2 and TMT/IRIS simulations were at the same exposure time and observational parameters.

\begin{figure*}
  \centering
  \includegraphics[width=1.\linewidth]{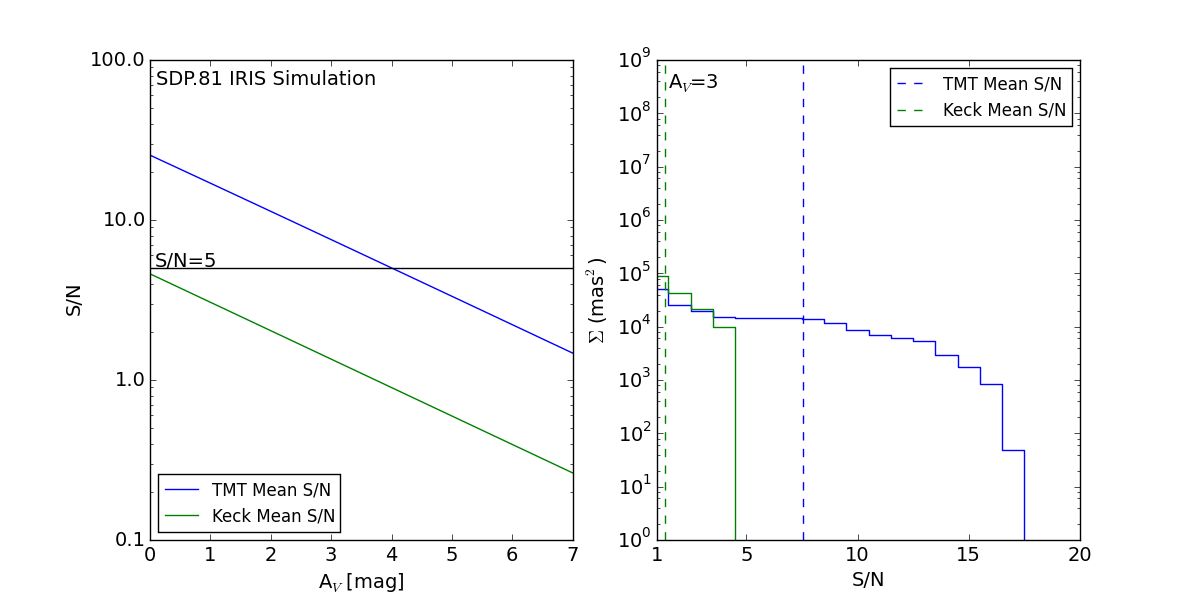}
  \caption{Signal-to-noise comparison between Keck/NIRC2 and TMT/IRIS from the simulation in Figure \ref{fig:sdp81}. (LEFT)  S/N of the simulations for a given A$_{V}$. IRIS+TMT will be able to perform essential imaging studies of heavily obscured ($\sim$A$_{v}$=4) high-redshift galaxies. (RIGHT) Number of pixels above a certain SNR at a dust obscuration of A$_{V}$=3. The median SNR for IRIS+TMT is about 7, whereas with Keck the maximum well below 2. TMT+IRIS will be able to resolve the stellar continuum in these dusty galaxies over lower surface brightness features with significant spatial coverage.}
  \label{fig:sdp81_s2n}
\end{figure*}

These simulations are a great example of how IRIS will be able to detect the stellar continuum for many highly dust attenuated (Av $>$2) gravitationally-lensed sources. Compared to current ground-based observations, IRIS will be able to measure lower surface brightness features at even greater S/N, as shown in Figure \ref{fig:sdp81_s2n}. Such impeccable observations of high-redshift gravitationally-lensed galaxies will yield essential constraints of resolved stellar masses, dust attenuations, star formation rates, metallicities, and kinematics.

\section{SUMMARY }
\label{sec:summary}  

TMT with IRIS and NFIRAOS will offer unparalleled abilities in \textit{many} diverse areas of astronomy. IRIS's capabilities of diffraction-limited sampling (4mas plate scales), moderate spectral resolving power (R=4000 - 8000) with an IFS, high astrometric accuracy (30 $\mu$as) with imaging, superb image quality from NFIRAOS, and unique parallel imaging-and-IFS mode, will be extraordinary even compared to future astronomical missions on the horizon, like JWST and LSST. Our team has developed an imaging and IFS data simulator for IRIS+NFIRAOS+TMT to explore a range of science cases. We have presented the most recent science cases that will truly exploit the diffraction-limited capabilities of both the imager and IFS, including studies of Pluto and Charon, the moon Io, spectroscopy of Galactic Center stars,  AGN torii, and individual star forming regions in gravitationally-lensed high-redshift galaxies. Our team will continue to use the data simulator to explore science cases and investigate instrument requirements for IRIS during the final design phase starting in 2017.

\acknowledgments 
 
The TMT Project gratefully acknowledges the support of the TMT collaborating institutions. They are the California Institute of Technology, the University of California, the National Astronomical Observatory of Japan, the National Astronomical Observatories of China and their consortium partners, the Department of Science and Technology of India and their supported institutes, and the National Research Council of Canada. This work was supported as well by the Gordon and Betty Moore Foundation, the Canada Foundation for Innovation, the Ontario Ministry of Research and Innovation, the Natural Sciences and Engineering Research Council of Canada, the British Columbia Knowledge Development Fund, the Association of Canadian Universities for Research in Astronomy (ACURA) , the Association of Universities for Research in Astronomy (AURA), the U.S. National Science Foundation, the National Institutes of Natural Sciences of Japan, and the Department of Atomic Energy of India.

\bibliography{report} 
\bibliographystyle{spiebib} 

\end{document}